Short Paper*

# A Tech Hybrid-Recommendation Engine and Personalized Notification: An integrated tool to assist users through Recommendations (Project ATHENA)

Lordjette Leigh M. Lecaros
Graduate School, Institute of Computer Science, College of Arts and Sciences
University of the Philippine Los Baños, Philippines
lmlecaros@up.edu.ph, lordjette.leigh@gmail.com
(corresponding author)

Concepcion L. Khan
Institute of Computer Science, College of Arts and Sciences
University of the Philippine Los Baños, Philippines
clkhan@up.edu.ph

## Abstract

*Purpose* – Project ATHENA aims to develop an application to address information overload, primarily focused on Recommendation Systems (RSs) with the personalization and user experience design of a modern system.

*Method* – Two machine learning (ML) algorithms were used: (1) TF-IDF for Content-based filtering (CBF); (2) Classification with Matrix Factorization- Singular Value Decomposition

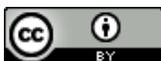

(SVD) applied with Collaborative filtering (CF) and mean (normalization) for prediction accuracy of the CF. Data sampling in academic Research and Development (R&D) of Philippine Council for Agriculture, Aquatic, and Natural Resources Research and Development (PCAARRD) e-Library and Project SARAI publications plus simulated data used as training sets to generate a recommendation of items that uses the three RS filtering (CF, CBF, and personalized version of item recommendations). Series of Testing and TAM performed and discussed.

*Results* – Findings allow users to engage in online information and quickly evaluate retrieved items produced by the application. Compatibility-testing (CoT) shows the application is compatible with all major browsers and mobile-friendly. Performance-testing (PT) recommended v-parameter specs and TAM evaluations results indicate strongly associated with overall positive feedback, thoroughly enough to address the information-overload problem as the core of the paper.

*Conclusion* – A modular architecture presented addressing the information overload, primarily focused on RSs with the personalization and design of modern systems. Developers utilized Two ML algorithms and prototyped a simplified version of the architecture. Series of testing (CoT and PT) and evaluations with TAM were performed and discussed. Project ATHENA added a UX feature design of a modern system.

*Recommendations* – High-end hardware specs of v-Parameter are recommended, at least 8-cores of vCPU with 16GiB of memory and sufficient driver-type size to run the model and End-to-end jobs execution for continuously maintaining the application.

*Research Implications* – Future Developers must use/integrate a large dataset. Other ML approaches can expand Hybrid-RS better. Implicit data and additional filtering methods may enhance the application in the future.

## INTRODUCTION

With the rapid spread of online information within the past several years, recommendation systems (RSs) are increasingly gaining momentum (Figure 1). They are presently a salient part of a modern system and render illuminating aid for users in searching sort of items and other services.

As an effective information filtering and recommendation tool for online information, RS has been broadly utilized in various areas: e-Commerce (e.g., Alibaba, Amazon, eBay, Lazada, and Shopee), movies (e.g., Netflix, Disney+, Apple TV, Movielens, Hulu, and Amazon Prime Video), music and video (e.g., Pandora, Spotify, YouTube, TikTok, and Last.fm), news (e.g., BBC News, Google News), Travel and Leisure (Expedia, Agoda), and other online platforms (More illustrations appear in Figure 1) (Afoudi et al., 2019; Agoda, 2021; Desai, n.d.; Fayyaz et al., 2020; Forbes Insights, 2020; Roettgers, 2019; Thann-Tuyen Tran, 2019; Wang et al., 2018; Xu et al., 2021; Yue, Wang, Zhang, & Liu, 2021).

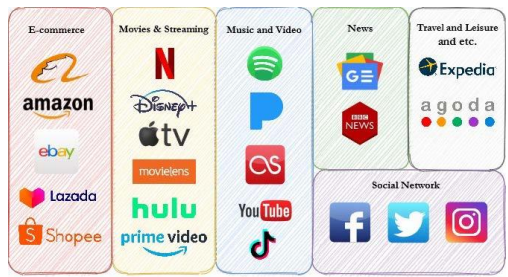

*Figure 1.* Recommender Systems utilized in various areas

However, filtering through all information and taking only the salient aspects can be challenging for everyday users. Searching through all options can be overwhelming; this can cause an overburden of data, especially for a surplus volume of online information has grown progressively troublesome to get relevant information and adequately evaluate retrieved items for making the right decisions. Non-personalization of items at any level can overlap any suggestions unrelated to the user's aspects of interest and without user experience of a modern system resulting in a pale of the applications.

RS features to personalize suggestions effectively by utilizing RS filtering methods to understand the user requirements from users' activity and data (Yue, Wang, Zhang, & Liu, 2021). At present, recommender system filtering can be generally cleaved into three-part, collaborative filtering (CF), content-based filtering, and hybrid RS filtering (combined filtering). Project ATHENA aims to develop a hybrid recommendation engine and personalized notification to address information overload, primarily focused on RSs for items with personalization and user experience design of a modern system, an application to assist users in searching sort of items thru recommendations.

Developers used a data sampling in academic Research and Development (R&D) of the e-PCAARRD library and Project SARAI open data to integrate the application with an online digital platform, a publication of items with a specific community (categories). Project ATHENA intends to engage researchers or users to the system by gaining interest in this online information and receiving a recommendation of items. This paper aims explicitly to apply the Two ML algorithms approach for Hybrid RS in the combination of CBF, CF, and personalized version of items and, depending on notification information (user profile and content information of items), user activity data (search, views, likes), item preferences, and delivery schedule to receive a recommendation of the item(s) via email notification. This paper also specifically aims to test the Compatibility and Performance of the application and evaluate it with Technology Acceptance Model (TAM).

## LITERATURE REVIEW

RS(s) have etched a spot for the masses through sturdy influence in various areas and have great success utilizing these systems nowadays (Yue, Wang, Zhang, & Liu, 2021). Fararni, K. A. et al. (2021) developed a Hybrid Recommender System for Tourism based on big data and AI. The study also aims to gain more visitor experience by recommending relevant items and helping visitors to personalize trips. The said system featured to go beyond a list of recommended tourist attractions and suggest an offer to a multi-day visit where users can choose diversified tourist spots or resources (activities, monuments, hotels, pool, and so on). Their architecture used innovative hybrid filtering, a combination of content-based items, collaborative filtering, and demographic filtering to identify user preferences and interests of a tourist in the selection process. Authors proved that their study solves the overcome information-overload problem triggered in the Tourism portal. Project ATHENA used hybrid filtering, the same with Fararni et al. (2021), a type of RS filtering in a combination of CF and CBF, to recommend an item with personalization and receive an email notification recommendation. The technique was based on recent user activity and search but limited in demographic filtering.

Moreover, a Hybrid Recommender System for Improving Automatic Playlist Continuation was studied and developed by Gatzioura et al. (2021). The said approach has proven an excellent help for Music listeners to find similar items and suggest other playlists related to the recent activities. The results stated that RS for playlist continuation had identified the desired characteristic to form a pleasant and satisfactory outcome for listeners. Therefore, Project ATHENA aims to recommend similar items with combined filtering and methods depending on users' recent activity on items such as likes, searches, and views.

Chakraborty (2018) developed and applied Data Extraction and Integration for Scholar Recommendation System. The author focused on academic recommendation, a framework to recommend the list of topics, contents, and areas of interest for scholars/professors, followed by algorithms to classify scholars with similar research interests. Results show that the author made relevant recommendations and confirmed that the framework is reliable and easy to use. Developers used data sampling records of Project SARAI and the e-PCAARRD library of Philippine Council for Agriculture, Aquatic, and Natural Resources Research and Development of the Department of Science and Technology (DOST-PCAARRD) that contains academic research and other publications. In addition, developers utilized CBF with meta-data available by its content and CF, a collaboration of users-correspond to items, and personalization at any level of the item in recommendations.

## METHODOLOGY

The Developers design a modular architecture of Project ATHENA, a Hybrid Recommendation System as an engine with filtering methods depending on: (a) Notification Information – user profile (name of the user, email address), content information of an item (title, publication date, descriptions, etc.); (2) User activity– user's recent activity search, user likes, and views to the item; (3) User preferences – The personalization of a user to the items in recommendations at any level to choose, a category of interest such as communities (banana, tomato, rice, etc.) or type of materials (Book, serial, thesis, etc.); (4) Delivery schedule– user personalization to schedule the notification, receive the recommending items and activate/deactivation setup. Developers used the said information for the Hybrid Recommendation System designed in the combination of Content-Based Filtering (CBF), Collaborative Filtering (CF), and personalized item (Figure 2).

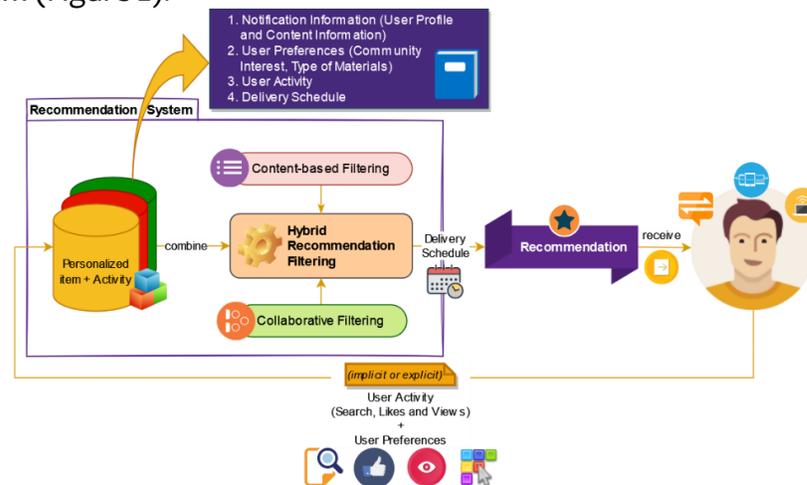

*Figure 2.* The Architecture of Project ATHENA

Content-based filtering (CBF), a pre-tagging to recommend an item with the same properties, this approach used to provide sufficient information and filtering in RS with item availability and based on similarities relevant to the user and the item. Collaborative filtering (CF), this approach assumes that users who have viewed or liked an item(s) recently will also consider and want the same in the future. Developers used this filtering method to build the RS that varies on the recent activity of users (searches, likes, views on the item). These activities seek the association between the users and the items. The Developers used the CF approach to predict the item scoring/ranking for the item that the user may be interested in, and the system will recommend this item. Developers utilized a *Model-based CF* approach to lessen the memory occupied by the CF and boost the application's performance, provide a rapid training speed, and achieve most cases in the accuracy of CF recommendation. MF is applied to discover underlying interactions between different entities and predict scores or ratings for CF to recommend items to users. Figure 3 shows the interaction between users and items in Matrix Factorization.

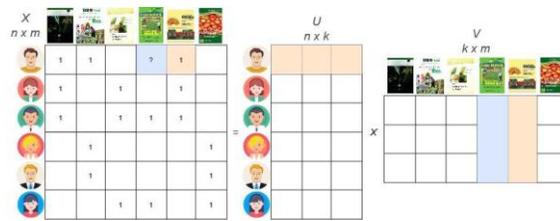

*Figure 3.* Illustration of matrix factorization

*Singular value decomposition (SVD)* is the first thing heard about MF-based algorithms due to their successful techniques and comprehensive extent utilization in the field of Mathematics in the long run. The prototype implementation utilized a classification model as MF, specifically the "SVD Model," to support the CF approach in this paper. Due to its fast-training speed and fewer memory consumptions, it obtains most cases in the accuracy of the retrieved item in the recommendation.

The Developers added other filtering methods to personalize recommendations based on user inclinations. The Developers designed this assumption as user interactions to customize and receive recommendation output through an email notification. The **personalized-item version** includes user preferences based on interest, delivery schedule, activation, and deactivation in receiving a personalized recommendation(s).

On the other hand, the Developers combined the advantages of CBF, CF, and personalized items to process different kinds of RS filtering with user preferences and the assumptions of similar interest to improve the prediction accuracy of the recommendation system. The use of combined RS filtering or hybrid-RS addresses the cold start problem, an item with insufficient information (metadata, or properties) available, which RS does not perform optimally. With RS combined filtering, this solves the data sparsity problem resulting from the fact that the users only intend to view or like limited items, so this paper used a *hybrid-RS* approach to provide a recommendation, personalization, and design of a modern system. The Developers used Project SARAI (Smarter Approaches to Reinvigorate Agriculture as an Industry in the Philippines) open datasets and the e-PCAARRD library of DOST-PCAARRD as data sampling containing as much needed information as possible. The Developers used the other synthetic data available and related only to crops commodities to collect the information on the context of this paper, including publications of items, books, thesis & dissertation, serials, non-prints, vertical files, inventory projects, technical reports, re-prints, analytics, journals, articles, posters and includes other information like user' full name, legit and synthetic email address as a training dataset.

The Developers picked a model to build the application. In CF, the Developers used a pre-built SVD model of SciPy, a library for scientific computing and machine learning. Since the idea of CF is about the user-item scoring matrix, the use of a Matrix-Factorization (MF) based algorithm was applied. The said library is well suited in developing a CF method to process the Classification of data using the said SVD, a type of MF. The formula produced by this model illustrates in equation 1, which decomposes a matrix $R$ into the best lower rank approximation of the original matrix $R$ (See Equation 1).

$$R = U\Sigma V^t \qquad \text{Equation 1}$$

Where $U$ and $V$ are orthogonal matrix (real square matrix) with orthonormal eigenvectors and $\Sigma$ are the diagonal matrix of singular values (actual weights), the matrix factorized to as shown in equation 2. The Developers intend to capture the underlying features or hidden-latent factors from users and items matrices by normalizing the scores and using the said model. The scoring predicts from the inner product of these two-factor matrices that recommendation output depends on the score varies in the score matrix populated by this model.

$$\underset{m \times n}{\mathcal{R}\begin{pmatrix} \alpha_{11} & \alpha_{12} & \alpha_{1n} \\ & \ddots & \\ \alpha_{m1} & & \alpha_{mn} \end{pmatrix}} \approx \underset{m \times m}{\mathcal{U}\begin{pmatrix} \mu_{11} & & \mu_{m1} \\ & \ddots & \\ \mu_{1m} & & \mu_{nm} \end{pmatrix}} \underset{m \times n}{\overset{\Sigma}{\begin{pmatrix} \sigma_1 & & & 0 \\ & \ddots & & \\ & & \sigma_r & \\ 0 & & & 0 \end{pmatrix}}} \underset{n \times n}{V^T\begin{pmatrix} v_{11} & & v_{1n} \\ & \ddots & \\ v_{n1} & & v_{nm} \end{pmatrix}} \qquad \text{Equation 2}$$

**Term Frequency-Inverse Document frequency (TF-IDF)**, a technique used in CBF, calculates the score for each document's description, word-by-word. The Developers used the TF-IDF pre-built model of Scikit Learn and trained the dataset to weigh a keyword in a record and assign its importance based on the number of times it appears in the document. Then, it simply put that the higher the TF-IDF score (weight), the rarer and more important the term, and vice versa (See Equation 3).

$$W_{x,\,y} = tf_{x,\,y} \times log\left(\frac{N}{df_x}\right) \qquad \text{Equation 3}$$

Moreover, the Developers designed an End-to-End application architecture of Project ATHENA. There are 4 phases in the architecture. *User Interface* - It mainly focuses on the front-end application. Users can view a list of items and recommendations, wherein project ATHENA integrates to an online digital platform. This phase also includes the personalization of recommendations to be received thru email notification. Developers used Flask, a micro web framework, as the front-end of this application. The phase also includes users-item interaction in which data collections on user activities like Search, Likes, Views of users are involved. *Data Ingestion and Processing* are collected using the Apache Spark library and Hadoop MapReduce to ingest various sources from User Activity Service, User Preferences Service, Notification Information Service, and Delivery Schedule Service. Once ingestion happens, data cleaning, formatting, and transformation are done to this phase and stored into the Delta Lake or Parquet as an alternative with Spark API and Hive. The Developers used physical and transactional tables as services from PostgreSQL DB and Elasticsearch collections data sources for the record statistics. After pre-processing of data, this trained and prepared for *Recommender Model* - used in this application. Hybrid RS implemented to this phase contains Python programming language with library and tools such as Pandas, Scikit-learn, SciPy, NumPy, Apache Spark,

JupyterLab, Delta Lake, or Parquet as an alternative to Delta Lake, these all used in the Machine Learning. The phase processed the Hybrid RS approach with three combined RS filtering methods: (a) Collaborative Filtering; (b) Content Filtering; and (c) personalized items that ingest and calculate the scores and mean to normalize the accuracy of results, then prepare a recommendation of items. **Store Data and Preparations**: In this phase, data with calculated scores and output persists/stores in PostgreSQL with Apache Spark library. Data ingestions and processing of data sources from various services are collected to prepare the data in joined with scores and filters the output to Hybrid Recommendation System. The Hybrid Recommendation System service, including the personalized version of item(s), consolidates all these to fetch the recommendation to the users' User Interface. Figure 4 shows the overall prototype implementation of how the Hybrid RS and personalized notification assembled to interact with each process from an end-to-end solution.

The main technological tools and programming languages used to build the application used Python, Flask web framework for UI, Elasticsearch, PostgreSQL, Spark, Jupyter, Google Charts for visualizations, Gitlab, Docker, and Google Cloud for deployment. In development, the Developers used a Lenovo ThinkPad E570 Laptop with 16GB of RAM and 50GB of SDD to develop the application with an Operating System of Ubuntu.

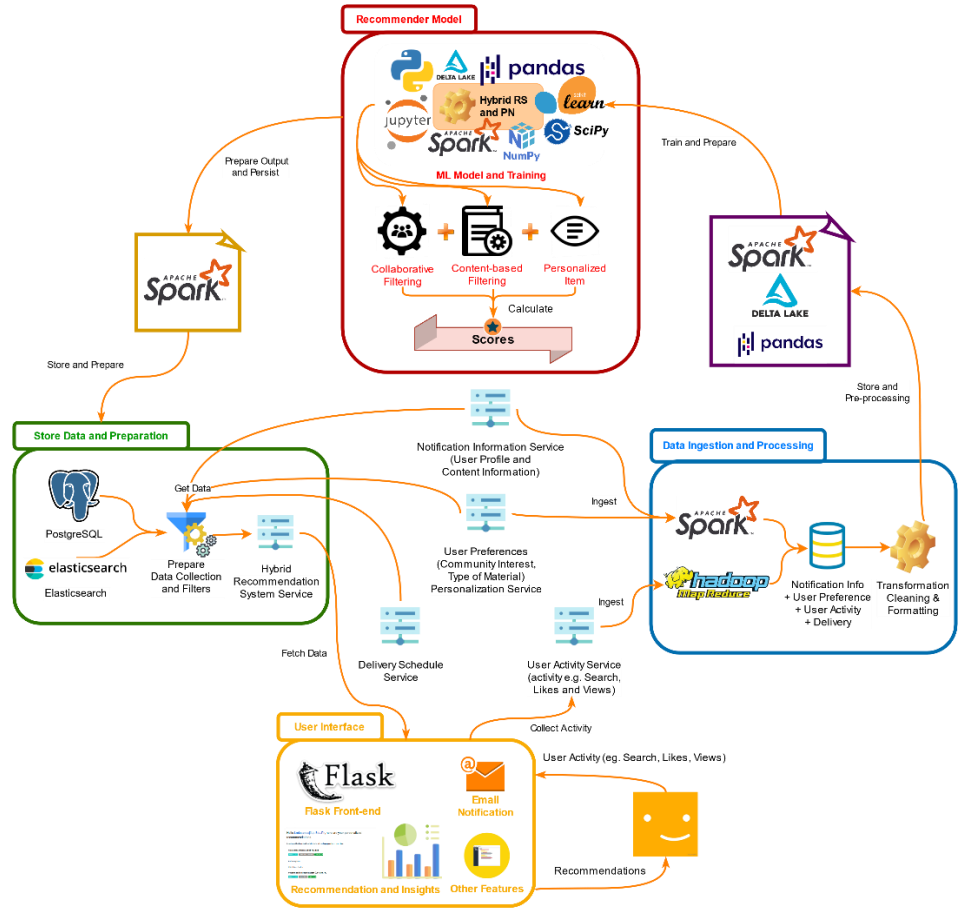

*Figure 4.* End-to-End Application Architecture

In addition, Developers evaluate and compare the ability of the Hybrid RS approach versus a single RS filtering (CF & CBF) and algorithm using the decision-based metrics. The 20% of datasets used in the training part. Precision, Recall, and F-measures utilized to identify the correct predictions from those incorrect predictions. Precision describes the number of selected items in relevant, and Recall signifies the number of selected items (Equation 4).

$$Precision = (\#RelevantRecommendations)/(\#RecommendedItems)$$
$$Recall = (\#RelevantRecommendations)/(\#PossibleRecommendedItems)$$
$$F-measure = \frac{2 \cdot Precision \cdot Recall}{Precision + Recall}$$

*Equation 4*

In addition, the developers led a series of testing with Compatibility Testing (CoT) to identify browser compatibility, operating system compatibility, and mobile browsing compatibility. Developers tested the application with Performance testing (PT), specifically endurance testing, to check the speed and system's behavior under a specific load/and continuous expected load. Parameters such as memory and CPU utilization are monitored in this PT to detect memory leaks, the rapidness of the application and to recommend the best solutions to utilize the application with v-parameters. In evaluation,

TAM was used to assess the application wherein an individual's behavioral intention to use guided a specific application by perceived usefulness and ease of use. The perceived usefulness factor of the application-defined extent to which users believe that using the system enhances their performance with added user experience design of a modern system.

**RESULTS AND DISCUSSION**

The application integrates into an online digital platform. Data sampling in academic Research and Development (R&D), particularly Project SARAI and e-PCAARRD library of DOST-PCAARRD publications, was used to build and test the said application. The page contains items (e.g., books, thesis & dissertation, serials, etc.). Each item includes one or more communities of categories: banana, cacao, coconut, coffee, corn, rice, soybean, sugarcane, tomato, and others (technology articles, announcements, etc.). Moreover, the Developers had designed a modern system user experience (UX) functionality. This RS(s) helps users search for items and provides suggestions relevant to a particular search, viewed, or likes.

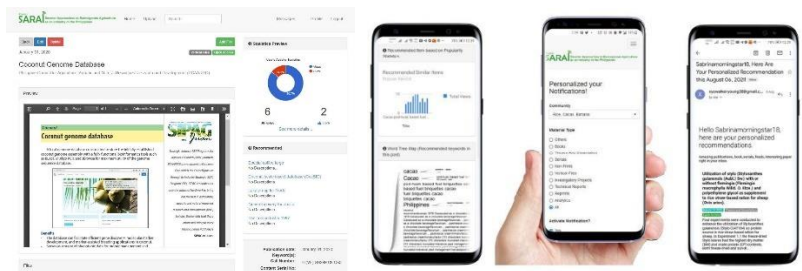

*Figure* 5. Application Interface

Figure 5 shows the recommendations for a particular item displayed in the in-lined text panel on the item page. The user can view these item recommendations, including the title and short description of these recommendations. Users can also view a statistics chart, personalized notification, and receive via email notification. Those recommendations produced by the hybrid RS application generate a fine-tuned relevant item to a specific publication and the user.

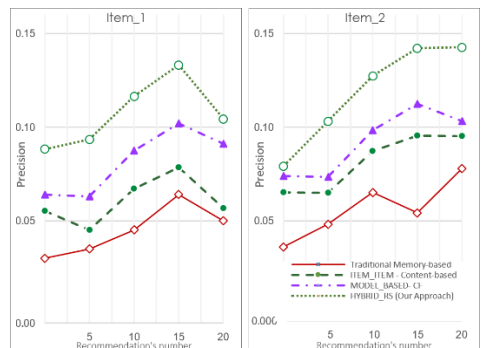

*Figure* 6. Recommendation's Precision Evaluation Results

The Developers evaluates and compares the Hybrid RS approach to various RS filters (CF and CBF) and algorithms using the decision-based metrics of Precision and Recall with F-measures. Findings found that hybrid RS shows to outperform the compared ones, this endorses the Developers' initial assumption that a hybrid RS prototyped for a particular problem that would be able to achieve improved than the usual RS techniques and resulting from the fact that the users only intend to view or like limited items, and while having a better trade-off with filtering beyond accuracy, in all the used training set. Results also correspond to plots in line with each evolution pattern. The higher, the better performance, and the lower inclination signifies lesser/failed to capture the recommendation of items (See Figure 6).

The Developers performed a series of testing to check the application's compatibility with specific browsers such as Chrome, Firefox, Edge, Opera, Safari, and new browsers like Brave and Vivaldi. The test was done and has achieved that the application is supported and compatible with the said browsers. In addition, Developers conducted OS compatibility. A minimum requirement is required to run the application in a Windows OS machine, such as drivers (matter of deployment choice for the model). While in some Linux-based OS, specifically Debian and Ubuntu, the application is compatible to run the required libraries and tools needed to deploy the said application.

Furthermore, developers tested the application in Smartphones to identify the mobile browsing compatibility as a matter of choice, which detects that the application is Mobile-Friendly. Developers also performed a PT, specifically endurance testing, to discover the application's performance, speed under sustained use, specific load, or continuous expected load. The application was tested and evaluated with the following virtual parameters (v-Parameters) of memory (driver and memory types) and v-CPU. The test was monitored and have identified the rapidness of the job execution per load with parameters run on specific VM Cores. Table 1 shows the following v-Parameters tested to compare the rapidness of the End-to-end job's implementation and Model execution of the application.

Table 1. v-Parameters

| vParameters | Driver (GB) | Memory (GB) | vCPU(Core) | Nodes |
|---|---|---|---|---|
| V1R1M8C4 | 8 | 8 | 4 | 1 |
| V2R1M14C4 | 14 | 14 | 4 | 1 |
| V3R1M16C8 | 16 | 16 | 8 | 1 |
| V4R1M28C4 | 28 | 28 | 4 | 1 |
| V5R1M28C8 | 28 | 28 | 8 | 1 |
| V6R1M32C8 | 32 | 32 | 8 | 1 |
| V7R1M32C16 | 32 | 32 | 16 | 1 |
| V8R1M56C8 | 56 | 56 | 8 | 1 |

As a result of v-Parameters, the duration with the lowest score (in seconds) holds much higher recommended v-Parameters to run and utilize the application. Running the model shows that the fastest v-Parameter was V8R1M56C8 which has 56 GB of memory type, same with the driver type, and at least eight cores of vCPU. Still, a significant factor for the end-to-end job execution, it is recommended to use either V7R1M32C16, which has 32GB of memory, and 32GB in the driver type size with 16 Cores of v-CPU, or V3R1M16C8 (current model used) has 16GB of memory type, same with driver type, and eight cores of vCPU, which is rightly enough, compatible, and cheaper to run the application (see Figure 7).

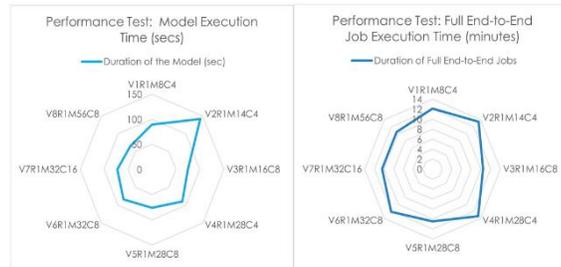

*Figure 7*. Performance Test Results

The Developers performed TAM to evaluate the application. A total of 25 evaluators participated in this assessment by answering the questionnaire via Google Forms and Google Sheets. Data collected was analyzed using Python based on the TAM evaluation through a quantitative test. Most results indicate that the application is practical, easy to use, and accepted to have a user experience design of a modern system (See Figure 8). Furthermore, the prototype and the evaluations received good feedback and have stated great ideas in recommendation given with user experience design of a modern system.

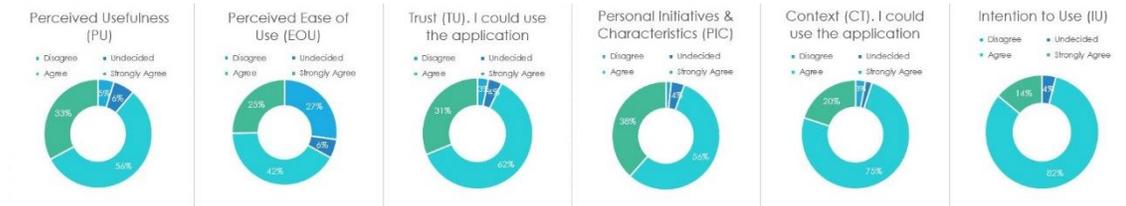

*Figure 8*. TAM Evaluation Results

## CONCLUSIONS AND RECOMMENDATIONS

In this paper, the Developers present a modular architecture of Project ATHENA, "**A**  **T**ech **H**ybrid-Recommendation **E**ngine and personalized **N**otification: **A**n integrated tool to assist users thru recommendations," which aims to address information overload. Two ML algorithms: (1) Term frequency-inverse document frequency (TD-IDF); and (2) Classification (Matrix Factorization using Singular Value Decomposition (SVD)) utilized in the said application. The TF-IDF algorithm applied to Content-based filtering (CBF), and Collaborative Filtering (CF) uses SVD with a mean (normalization)(Eding, 2019) that calculates the prediction accuracy of CF recommendation filtering. The Developers simplified the version of the architecture and discussed the preliminary prototype implementation. Project ATHENA generates and sends a recommendation of the item(s) produced by three (3) combinations of RS filtering (CF, CBF, and personalized version of item recommendation). It depends on notification information, user activity (search, likes, and views), inclusion to the RS filtering with user preferences, and delivery schedule to receive the recommendation of item(s) to be sent via email notification. In addition, the application has added UX design features of a modern system. Moreover, the application tested and performed a series of testing (Compatibility testing (CoT) and Performance Testing (PT)). The application appears compatible with all significant browsers and shows a mobile-friendly application. PT findings recommended three (3) v-Parameters V8R1M56C8, with 56 GB of memory and driver types with eight cores. Still, in a more excellent factor, at least for the end-to-end job execution, it is recommended to use either V7R1M32C16, which has 32 GB of memory type same with driver type and has 16 Cores of v-CPU, or V3R1M16C8 that has 16 GB of memory and 16 GB in the driver type size and has eight cores, which is cheaper but also fast to run the Recommender model and end-to-end jobs execution of the application, and continuously maintain as the core of the future efforts of this paper. TAM evaluations results received most of the positive feedback in the following criteria: Perceived Usefulness (PU); Perceived Ease of Use (EOU); Trust (TU); Personal Initiatives and Characteristics (PIC); Context (CT); Intention to Use (IU), given the satisfactions and features have stretched the possibility, scope, and objectives of this project application.

## IMPLICATIONS

Future Developers must use a large dataset, which can be helpful in the model. Other ML approaches can enhance Hybrid RS better. The use of implicit data in filtering may expand the recommendation as well. Project ATHENA needs further improvements such as (a) Scalability and integration to other platforms, and future maintenance helps to improve the applications. (b) Synonymy, Abbreviations, location-based, and demographic filters adding these to Hybrid RS filtering may improve the applications in the future. (c) Enhancement of the prototype to include unconsidered blocks in sending email notifications as well.

## ACKNOWLEDGEMENT

To the University of the Philippines Los Baños, the Graduate School, together with the Institute of Computer Science, to all the faculty and staff, thank you for letting me learn and improve through this cornucopia of knowledge and experience. All Words aren't enough to express how thankful I am to accomplish this Project. Hence, I am forever grateful to all of you.